\documentstyle[epsfig]{aipproc}

\newcommand{\be}{\begin{equation}}
\newcommand{\bfig}{\begin{figure}}
\newcommand{\ee}{\end{equation}}
\newcommand{\efig}{\end{figure}}
\newcommand{\bi}{\begin{itemize}}
\newcommand{\ei}{\end{itemize}}
\newcommand{\bear}{\begin{eqnarray}}
\newcommand{\eear}{\end{eqnarray}}
\newcommand{\ba}{\begin{array}}

\newcommand{\ea}{\end{array}}

\begin{document}
\title{Scaling and Multi-scaling in Financial Markets}

\author{Giulia Iori}
\address{Department of Mathematics,\\ 
King's College London,\\
 The Strand, London WC2R 2LS, UK.}
\maketitle

\begin{abstract}

This paper 
reviews some of the phenomenological  models which have been 
introduced to incorporate the scaling properties of financial
data.  It also  illustrates  a microscopic
model, based on  heterogeneous interacting agents,
which provides a  possible  explanation for
the complex dynamics of  markets' returns.
Scaling and multi-scaling analysis performed on  the simulated data is in
good quantitative agreement with the empirical results.

\end{abstract}

\section{Introduction}

The    probability  distribution of returns 
of
 many market  indexes and exchange rates,
at  a given (but not too long) time scale  seems   consistent, 
 with   an asymptotic   power law decay $P(r) \sim r^{-(1 + \mu)}.$ 
A crossover between a Levy stable regime with  $\mu < 2$ 
\cite{Mantegna},  and a power law  with an exponent 
 $2 < \mu < 4$ \cite{Pagan} has been reported \cite{Gopikrishnan}.
Power law distributions are self-similar, therefore
the distribution of   returns  
$r_{\tau}(t) = \log (P(t + \tau)/P(t))$
 at different time scales 
$\tau$,
 rescaled by a lag-dependent factor $\eta(\tau)$, satisfies
\be
P(r_{\tau}) =  \frac{1}{\eta(\tau)} 
{\cal{F}}(\frac{r_{\tau}}{\eta(\tau)})
\ee
where $\cal{F}$$(u)$ is a time independent scaling function.
For self-affine function $\eta(\tau) = \tau ^H$ with $H=1/2$ for
a Gaussian process  and $H = 1/\mu$ for a Levy process.
If  scale consistency holds
 the time-scale at which the  process is observed
   becomes irrelevant.  
 The situation in financial markets 
 is more complicate  than this and one observes that  when $\tau$
 increases over a certain value 
(a few days for market indexes and a few weeks for individual stocks)
the scaling breaks down and the  
the shape of a Gaussian, as predicted by the
 random-walk hypothesis, is  recovered.
This implies that it is not possible to find a unique
 real number $H$ such that the
statistical properties of the rescaled  variables $r_\tau(t)/\tau^H$ do not
depend on $\tau$.
The notion of multiaffinity is introduced
to  characterize a stochastic process $X(t)$  which satisfies
\be
E(|X(t + \tau) - X(t)|^q)  = c(q)  \tau^{\phi_q}
\ee
where $\phi_q$ is the scaling function. Self-affine
processes are characterized by a  $\phi_q $ which  is linear
 and fully determined by its index $H$: $\phi_q = Hq$. 
Multi-affine processes instead are characterized by a non-linear 
scaling function $\phi_q$.
Anomalous scaling, or multiscaling,  appear in a wide class of phenomena where 
global dilatation invariance fails.
For example  intermittent behaviour in dynamical system, i.e. 
strong  time
dependence in the degree of chaoticity, 
is accompanied by 
anomalous scaling with respect to time dilatations in the trajectory
space.
If the non linear shape of the scaling exponents $\phi_q$ is 
a consequence of  intermittent behaviour
 multiaffinity  also involves  multifractality of an opportunely defined 
 probability measure.
A measure of the degree of intermittency could than  be provided
in terms of  an infinite set of exponents associated   to the  geometrical
structure of this  probability measure \cite{Calvet}.

Multiscaling in financial markets 
is an indication that returns, even if uncorrelated, 
are not independent stochastic
variables  and it reveals the presence of wild fluctuations.
The wilder the fluctuations, the
larger the difference of $\phi_q$ from a linear behaviour in $q$.
Indeed it is well known that while stock market returns are
uncorrelated,  the autocorrelation
function  of a measure of volatilty, such as absolute value of  returns, 
 is positive and slowly
decaying, indicating  long memory effects.
 This phenomenon, known in the literature as  volatility clustering
 implies temporal dependencies in the alternation of period of
 large price changes with period of smaller changes.  
Empirical analysis  
shows  that  the  decay of the  autocorrelations of absolute returns
 is hyperbolic 
over a large  range of time lags (from one day to one year),
 with an exponent  $0.1 < \gamma < 0.4$ \cite{Ding}.

Considerable  attention 
has been devoted in detecting 
 comovements of volatility with other economic
  variables in the attempt to 
interpret and capture the source of clustering effects in returns. 
In particular a big  effort has been devoted
to the analysis of correlations between  volatility of returns and 
trading volume.
Empirical evidence has been provided \cite{Pagan}
of a  positive cross correlation between  these two quantities. 
Furthermore  a  lack of regularity  has been  observed 
in  the trading or business time. Stochastic trading time models
\cite{Clark,Dacorogna,Scalas}  have also been 
proposed as a possible explanation for the 
emergence of  persistency in volatility.

\section{Phenomenological  Models}

A simple  model  proposed by Mandelbrot and van Ness \cite{Mandel_2} 
which can  generate  volatilty persistence is the 
 Fractional Brownian Motion. FBM is a random process with stationary, 
self-similar increments which grow locally at a rate $\tau^H$, 
where $H$ is the self-affinity exponent.
Even thought this model account for correlated volatility fluctuations, 
these are  generated through
dependent, and hence predictable, price  increments with negative
autocorrelation when $0<H<1/2$ 
and positive correlation when $1/2<H<1$.

An earlier model  introduced by Mandelbrot \cite{Mandel_3}, which has
become very popular in the physics literature, is the 
 Levy Flight (LF) model. 
A LF is a random walk in which the step lenght is chosen 
from a Levy distribution.
Since Levy distributions are stable under convolution
 the LF process exhibits  exact  self-similarity. 
The exact scaling  of both FBM and LF  is not consistent though
 with the 
scaling  break down  observed in financial market data.
To overcome the many limitations of Levy flights
(not last the difficulties to deal with  a process
with a diverging  second moment) 
Truncated Levy flights (TLF)  have been introduced. 
A TLF  is a process 
 based on a truncated Levy stable distribution with a cut-off
 in its power law tail. 
 The truncation can be introduced as a sharp cut-off as in
 Mantegna and Stanley \cite{Mantegna} or,
 to preserve the infinitely divisible property
 of the pdf   through a smoother  exponential decay in the tail as in 
 Koponen \cite{Koponen}.
Because of the cut-off 
 the distribution is no longer  self-similar  when convoluted 
and has finite variance. Even though the  TLF pdf
   belongs to the basin of attraction of a 
Gaussian, it  converges to the  Gaussian 
very slowly and the process exhibits Levy scaling 
in a wide range of sampling intervals.
Nakao \cite{Nakao} has shown that TLFs a la Koponen  exhibit the 
simplest form of multiscaling, i.e. bi-fractality. Nonetheless his
results, i.e. $\phi_q = q/\alpha$ for $0 < q < \alpha$ and  
 $\phi_q = 1 $ for $q > \alpha$ are not consistent 
with the empirical findings.
Similar results 
 have been found by Chechkin \cite{Chechkin} 
 who analized  a finite sample of a simulated  ordinary  LF. 
Indeed, even though  Levy flights have stationary,
 statistically self-affine and
 stably distributed increments,
  the finiteness of the sample size violate sef-affinity giving
 rise to spurios multi-scaling. Consequently, while the
 moments of stable Levy distribution with 
$\alpha < 2$   diverge,  together with the q-th order
 structure function, at $q>\alpha$  both 
quantities are finite for finite sample size.
A rough estimate of finite sample effects gives \cite{Chechkin}
 a linear
 $\tau$-dependence of the q-th order structure function with an
 exponent  which does not depend on q at $q>\alpha$.
So both ordinary  or truncated Levy flights can generate multiaffinity
 but the shape of the scaling function $\phi_q$ is not consistent 
with the empirical one.

Few aspects remain unexplained by the TLF: first is that the TLF 
describes well only the central part 
 of  distribution of return  but not the far tails
which  decay with an exponent $\mu$ well outside the Levy regime. 
Second the crossover to the Gaussian regime occurs at much larger
 times than the one
expected from the TLF
and, finally,  it does not account for correlated variance 
fluctuations.

In the economics literature the modelling of financial time
 series developed in a quite 
different direction and leptokurtotic  distributions 
have been introduced through a  variance conditionally dependent
on its past (heteroschedasticity), as in the ARCH/GARCH models 
\cite{Bollerslev}.
GARCH models nonetheless, even if fat-tailed, 
  only achieve weak  memory effetcs,  manifest in
an  exponential (and not hyperbolic)
  decay of the volatility autocorrelation function. 
 The recently developed FIGARCH \cite{Baillie}
  process achives long memory still  mantaining
 the martingale property of asset returns. 
Nonetheless, unlike GARCH, FIGARCH does  not
converge to a  Gaussian process  over long sampling intervals and fails
 to describe the scaling properties of pdfs at different time horizons.

 As a generalization of ARCH models 
 Heteroskedastic  Levy Flight (HLF) processes have  been recently 
introduced.
Podobnik et al.  
\cite{Podobnik} show that due to the correlations in the variance the
 process generates power-law tails in the distribution of returns 
whose exponents can be controlled through the way the correlations in
 the variance are introduced. The crossover between the two power-law
 regimes ($\mu <2$ and $2<\mu<4$) can also be generated in these models. 
Santini \cite{Santini} moreover has  shown that in the HLF 
the Levy scaling of the pdf persists for times
 of order of magnitude larger than for uncorrelated variance fluctuations.
It would be interesting to analyze whether this model also  reproduces
 the multiscaling behaviour of financial data.

An alternative exactly soluble model that mimics the long range volatility 
correlations has been introduced by Bouchaud et al. \cite{Bouchaud}. Although
 their model is uni-affine by construction, it shows apparent multiscaling,
 in good agreement with empirical data,  
as a result of very long  transient effects, induced by the long range
 nature of the volatility correlations.

This scenario  seems to 
indicate that  multiscaling in financial markets is not 
a trivial effect of   the truncature in the 
 tails but is  a consequence of correlated variance
fluctuations.

Following a completely different  approach, an altenative model, 
the Multifractal Model of Asset Return (MMAR),  has been introduced 
by Mandelbrot \cite{Mandel_4}.  
Fluctuations in volatility are introduced in MMAR by a random trading time,
 generated as the cumulative ditribution function of a random 
multifractal measure. Trading time is  assumed 
 highly variable and contains long memory. Both these characteristics
 are passed on to the price process trough compounding. 
 Subordinate stochastic process 
have been extensively used in the economic literature
where either the trading volume
 \cite{Clark}  or the trading time \cite{Dacorogna,Scalas}
has been chosen  as the directing process. Note that 
in general the distributions of 
subordinated stochastic processes  do not possess scaling properties.
 In the MMAR model the return
 process is a compound process
$X(t) = B_H[\theta(t)]$
where $B_H(t)$ is a fractional Brownian Motion with self-affinity index $H$,
and $\theta(t)$ is a stochastistic trading time. In particular
 $\theta(t)$ is a multifractal process with continuous, non-drecreasing
 paths and stationary increments. If $B_H(t)$ is 
 a Brownian motion ($H$ = 1/2) without drift   the MMAR generates,
 together with multiscaling, uncorrelated increments and persistence
 in volatility. 

Aside the phenomelogical characterization of the scaling
 properties 
of financial data the microscopic origins of the complexity of
 financial markets  needs to be investigated. From a physicist point a 
view the market is a perfect example of a complex system, 
with a large number of
heterogenous agents interacting in an intricate  way. 
It is than tempting to describe the behaviour of  market's player
  using  models developed in the context of 
 the statistical mechanics of disordered systems, as it is proposed
in the following section.

\section  {A Microscopic  Model}

 It is not settled yet
whether the emergence of power law fluctuations and volatility clustering
observed in financial data 
is due to external factors, like the arrival of new information, or
to the inherent interaction among market players
and the trading process itself. 

Iori  \cite{Iori}  proposed a model where
 large fluctuations in  returns   arise  purely  from communication and
imitation among traders. The key element in the model is the
introduction of a trade friction (representing transactions costs)
 which, by  responding
 to price movements, creates a   feedback mechanism
 on  future trading and generates volatility clustering.
In \cite{Iori} (to which the reader is refered for more details
 and for references of alternative agents based models), 
 the market  consists of a  market maker plus a number of noise
traders. Traders buy from or sell to the market maker
and respond to a signal which incorporates idiosyncratic
preferences and  the influence of the traders  nearest to them.
Only one kind of stock is traded, whose price is set
by the market maker on the basis  of the   observed  order flow and the
 overall trading volume.

The model is studied through numerical simulations. It repoduces
 correctly the  scaling of 
the  distribution of returns, with a power law decay with  $\mu \sim 3$, 
at short time and a slow   convergences to a Gaussian distribution at 
larger time scales. Moreover the model generates volatility 
clustering and positive cross-correlation between volatility 
and trading volume. The autocorelation function of the volatility 
decays hyperbolically with an exponent $\gamma \sim 0.3$ consistent 
with empirical observations.

\begin{figure}[b]
\centerline{\epsfxsize 7cm\epsfbox{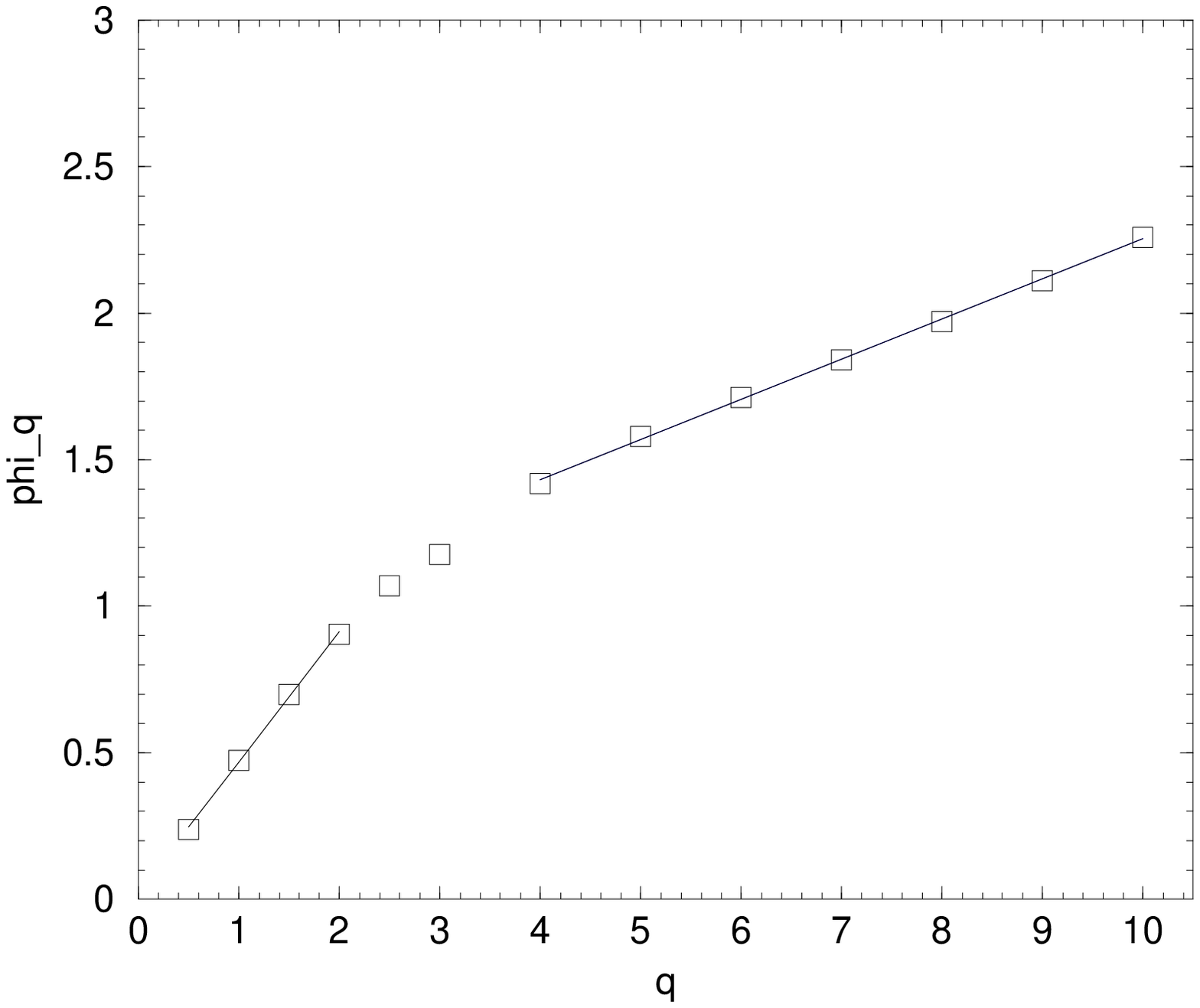} \hskip 1cm
\epsfxsize 7cm\epsfbox{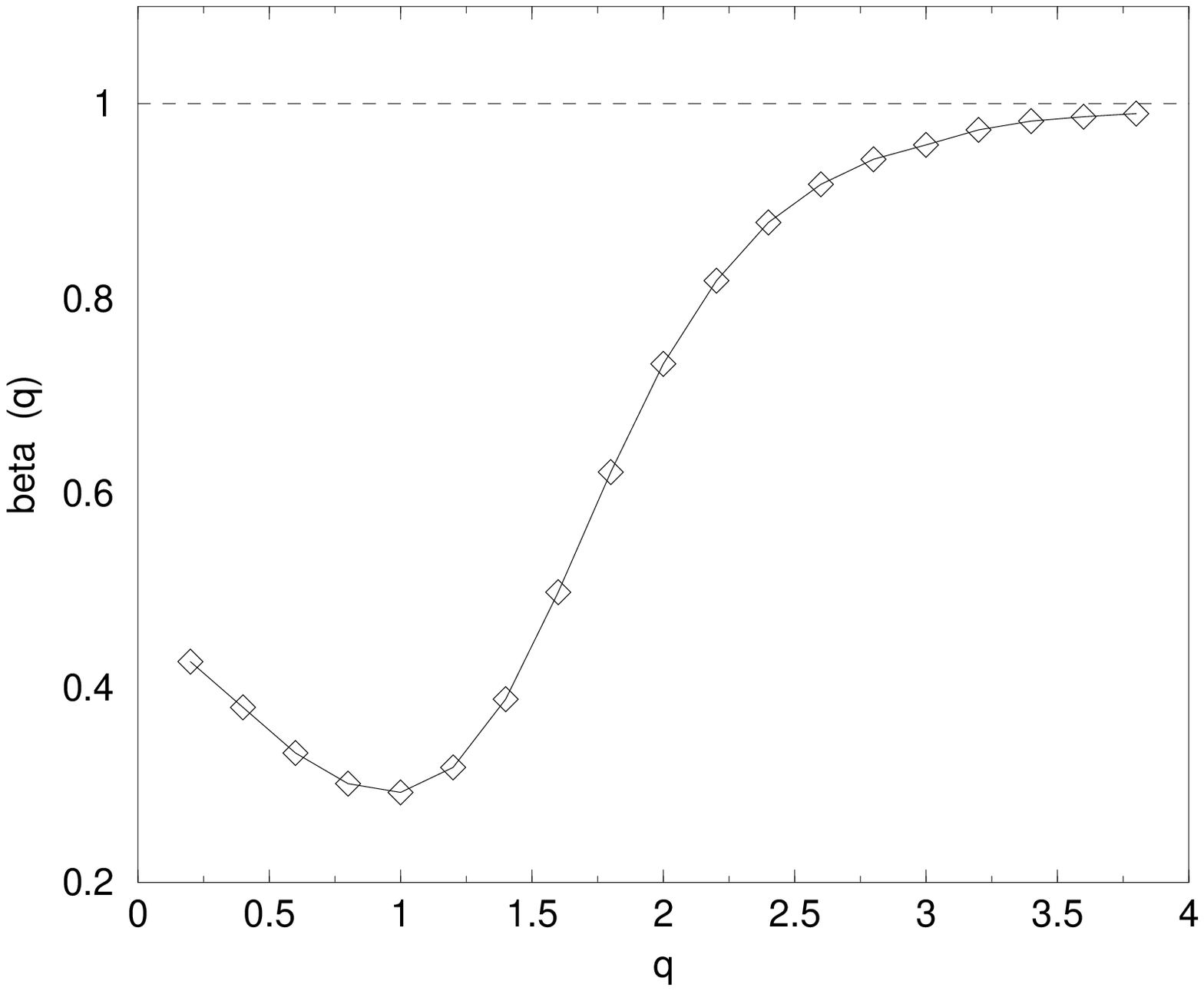} \hskip 1cm }
\vskip .2cm
  \caption[1]{ Exponent $\phi_q$ versus $q$  (left). The slope of the two regions
  are 0.47 at $q < 3$ and  0.14 at $q > 3$.
Scaling exponent $\beta(q)$
 as a function of $q$ (right). Anomalous
  scaling with  $\beta(q) < 1$ is shown.}
  \protect\label{FIG1}
\end{figure}

The  analysis of the moments of the distribution of the
simulated data also reveals multiscaling (fig.(1a)).
 By using a step-wise  linear
regression, the slopes in the structure function are estimated as  
 $0.47$ for $q < 3$   and  $0.14$ for $q > 3$.
These  results are in
good  agreement with the empirical analysis performed by Baviera
 et al. \cite{Baviera} on the $DM/US$
exchange rate quotes.

Baviera et al. \cite{Baviera} also detected
multiscaling in the  volatility autocorrelations
by  analyzing  the  generalized correlation functions: 
\be
C_q(L) =  <|r(t)|^q |r(t+L)|^q> -
 <|r(t)|^q><|r(t+L)|^q>.
\ee
If the absolute returns' serie $|r(t)|^q$ has long term memory
we would find
$C_q(L) \sim L ^{-\beta(q)}$
with $\beta(q) < 1$
while if $|r(t)|^q$ is an uncorrelated process   $\beta(q) =
1$. Multi-scaling would be  signaled by a non linear shape of $\beta(q)$.
The scaling exponent $\beta(q)$ measured from simulated data
is shown in fig.(1b) for $0 < q < 4$.
In analogy with the  NYSE index and the USD-DM exchange rate
\cite{Baviera,Pasquini},
 $\beta(q)$ is not a constant function of
$q$ revealing the presence of different anomalous scales.
The convergence of $\beta(q)$ to one reveals that large fluctuations are
practically independent. This observation might  justify
 the  convergence of the distribution of returns  to a Gaussian  even 
though  returns are not  independent random  variables.

This simple model can  reproduce  many of  the stylized 
facts of stock market returns  and has outlined a mechanism  
which  can explain  the emergence of the observed 
power-law fluctuations.

\end{document}